\documentclass[letterpaper]{article} 
\usepackage{aaai24}  
\usepackage{times}  
\usepackage{helvet}  
\usepackage{courier}  
\usepackage[hyphens]{url}  
\usepackage{graphicx} 
\usepackage{natbib}  
\usepackage{caption} 
\usepackage{algorithm}
\usepackage{algorithmic}
\usepackage{booktabs}
\usepackage{subcaption}
\usepackage{enumitem}
\usepackage{multirow}
\usepackage{amsfonts} 

\usepackage{newfloat}
\usepackage{listings}
\DeclareCaptionStyle{ruled}{labelfont=normalfont,labelsep=colon,strut=off} 
\floatstyle{ruled}
\newfloat{listing}{tb}{lst}{}
\floatname{listing}{Listing}

\usepackage{xcolor}

\usepackage{cleveref} 

\newtheorem{definition}{Definition}

\title{Scaling Lifelong Multi-Agent Path Finding to More Realistic Settings: \\ Research Challenges and Opportunities}

\author {
    He Jiang,
    Yulun Zhang,
    Rishi Veerapaneni,
    Jiaoyang Li
}
\affiliations {
    Carnegie Mellon University\\
    \{hejiangrivers,yulunzhang,vrishi,jiaoyangli\}@cmu.edu
}

\begin{document}

\maketitle

\begin{abstract}
Multi-Agent Path Finding (MAPF) is the problem of moving multiple agents from starts to goals without collisions. Lifelong MAPF (LMAPF) extends MAPF by continuously assigning new goals to agents. We present our winning approach to the 2023 League of Robot Runners LMAPF competition, which leads us to several interesting research challenges and future directions. In this paper, we outline three main research challenges. The first challenge is to search for high-quality LMAPF solutions within a limited planning time (e.g., 1s per step) for a large number of agents (e.g., 10,000) or extremely high agent density (e.g., 97.7\%). We present future directions such as developing more competitive rule-based and anytime MAPF algorithms and parallelizing state-of-the-art MAPF algorithms. The second challenge is to alleviate congestion and the effect of myopic behaviors in LMAPF algorithms. We present future directions, such as developing moving guidance and traffic rules to reduce congestion, incorporating future prediction and real-time search, and determining the optimal agent number. The third challenge is to bridge the gaps between the LMAPF models used in the literature and real-world applications. We present future directions, such as dealing with more realistic kinodynamic models, execution uncertainty, and evolving systems.

\end{abstract}

\section{Introduction}
In Multi-Agent Path Finding (MAPF)~\cite{SternSoCS19}, agents are each assigned a pair of start and goal locations, and algorithms attempt to find collision-free paths for them. MAPF has been widely studied due to its valuable applications in automated warehouses, traffic systems, advanced manufacturing, etc. However, many real-world systems continuously run for a very long time, where agents will be assigned new tasks. Such scenarios are better modeled by an extension of the MAPF problem, called Lifelong Multi-Agent Path Finding (LMAPF) \cite{MaAAAI17}. In recent years, many researchers have studied LMAPF and its variants \cite{LiuAAMAS19,salzman2020research,Li2021RHCR,DamaniRAL21,varambally2022mapf,zhangLayout23}, especially on its application in coordinating mobile robots in warehouses, which is nowadays a multi-billion industry. 

\begin{figure}[!t]
    \centering
    \begin{subfigure}[b]{0.265\linewidth}
        \includegraphics[width=1\linewidth]{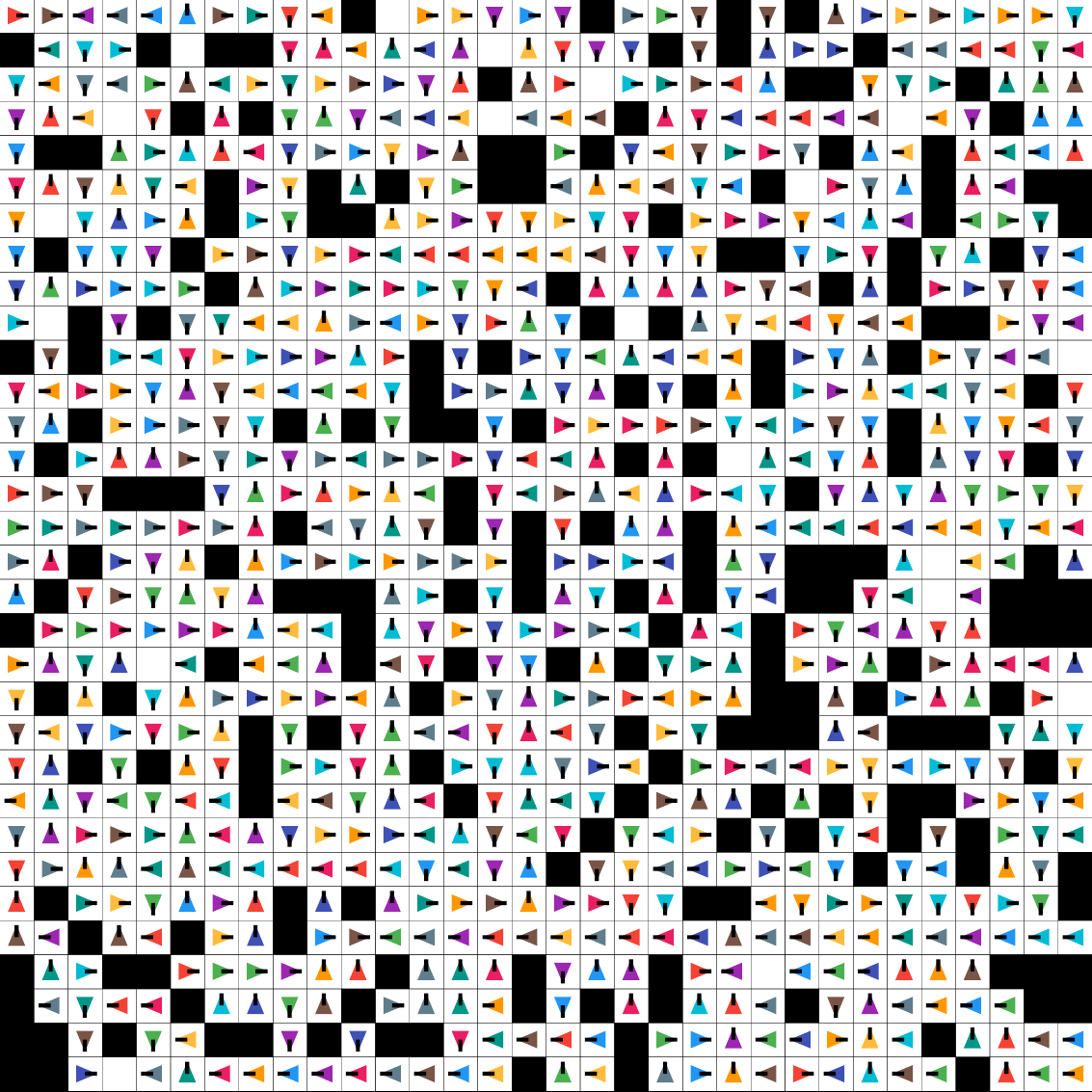}
        \caption{Random 800}
        \label{fig:front-fig-random}
    \end{subfigure}
    \hfill
    \begin{subfigure}[b]{0.44\linewidth}
        \includegraphics[width=1\linewidth]{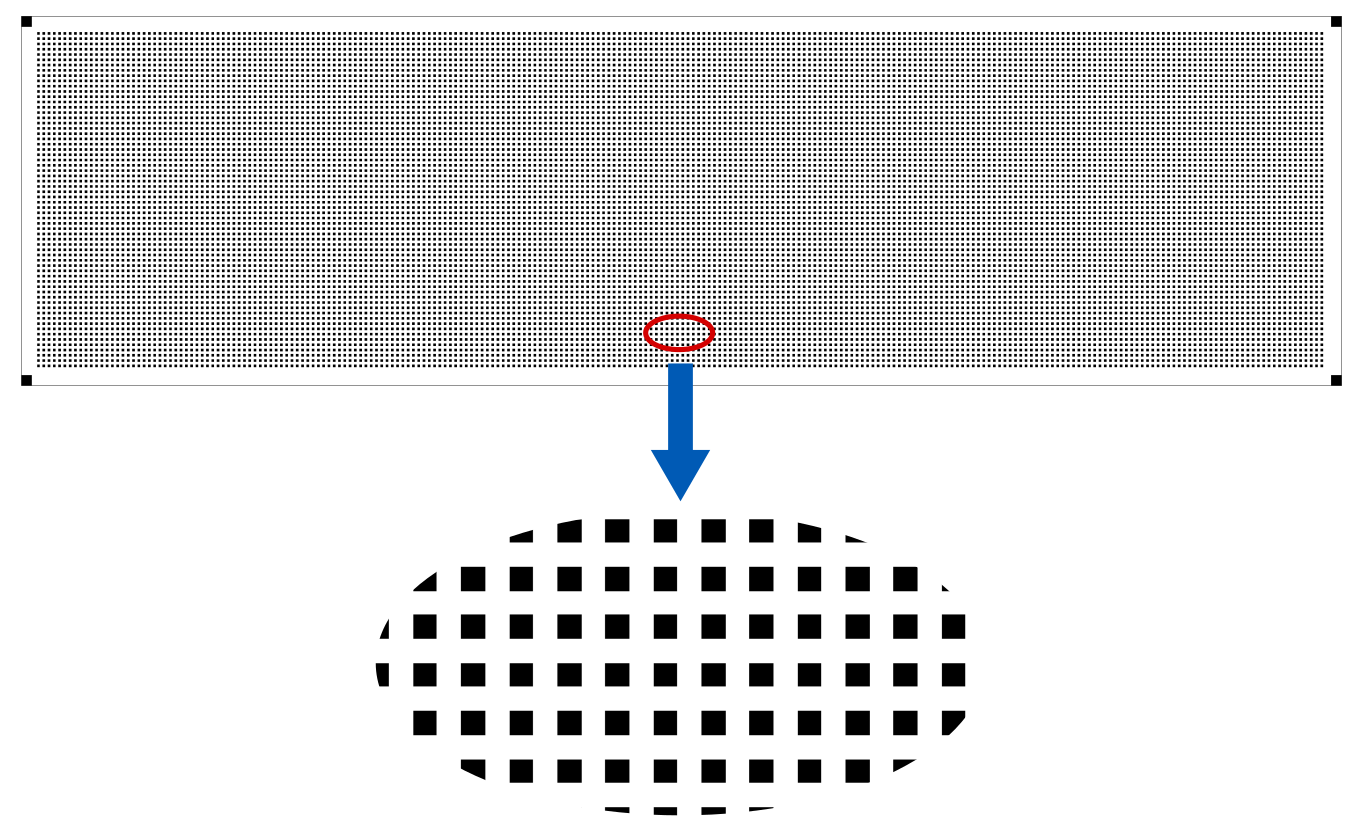}
        \caption{Sortation}
        \label{fig:front-fig-sortation}
    \end{subfigure}
    \hfill
    \begin{subfigure}[b]{0.265\linewidth}
      \centering
      \includegraphics[width=1\linewidth]{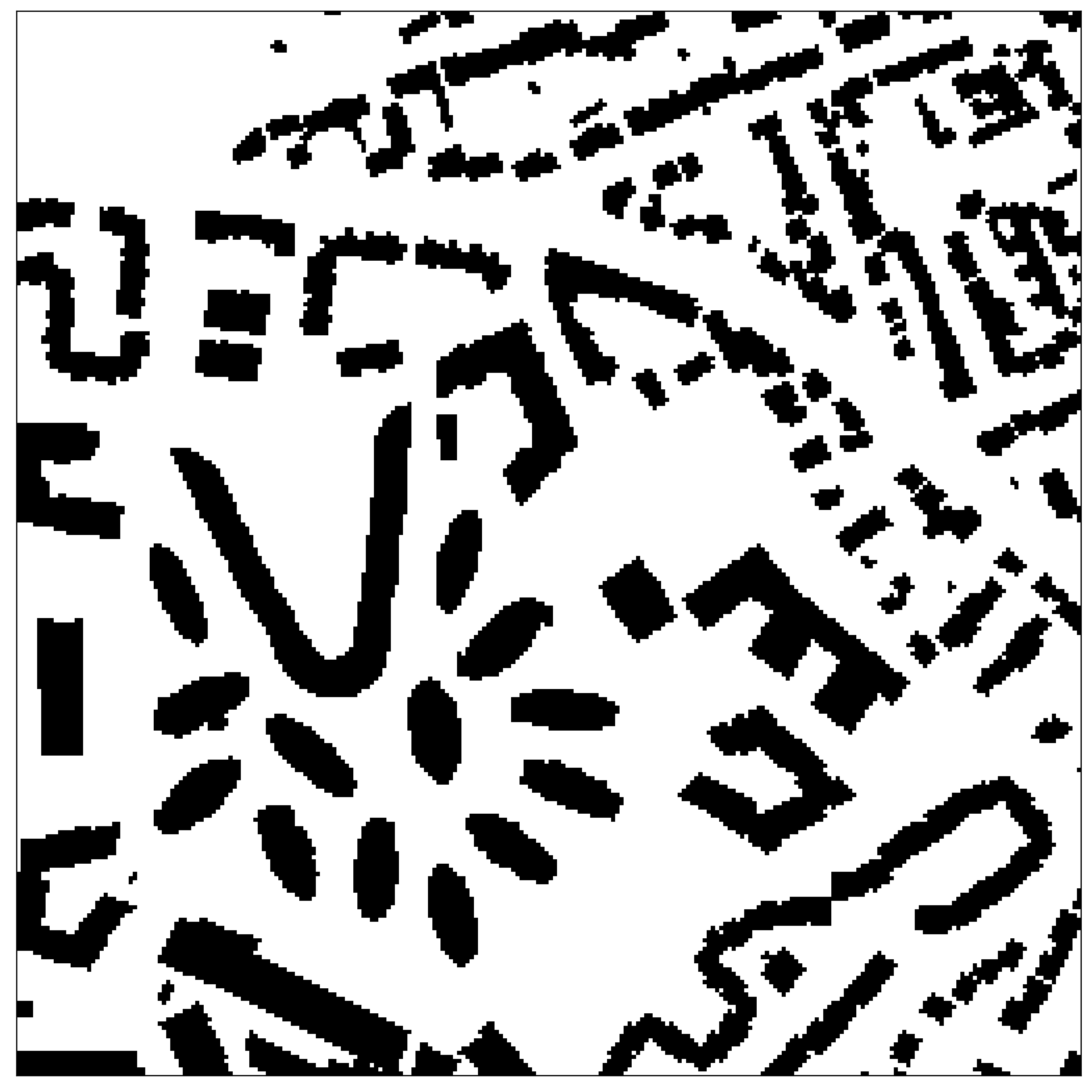}
      \caption{City}
      \label{fig:map-city}
    \end{subfigure}
    \caption{Three challenging maps in the LRR competition with white vertices and black obstacles. (a) is a random map of size 32 by 32 with very high agent density (800 agents out of 819 vertices
    ). The colorful triangles are agents, with short black lines indicating orientations. (b) is a large sortation center map of size 140 by 500 with 54,230 vertices and 10,000 agents. (c) is a city map representing a part of Paris. }
    \label{fig:front-fig}
\end{figure}

However, most MAPF problem instances studied in previous work are relatively simple. For example, in the widely used MAPF benchmark~\cite{SternSoCS19}, the agent density is no more than $50\%$, the number of agents is no more than $1,000$, and the planning time limit is usually set to 1-5 minutes. However, in real-world applications, automatic garage~\cite{guo2023valet} could have a near-100\% agent density, Amazon fulfillment centers could have more than 4,000 robots~\cite{Brown2022amazonrobot}, and cooperative autonomous driving requires real-time response. Only a few MAPF research, such as LNS2 \cite{li2022lns2} and LaCAM \cite{okumura2023lacam}, have tested with the more challenging instances. Similarly for LMAPF, only a few studies \cite{Li2021RHCR,DamaniRAL21} test with up to $1,000$ agents. One recent work~\cite{ChenAAAI24} scales the agent number to more than $10,000$ on large maps. 

Thus, an LMAPF competition, The League of Robot Runners (LRR),\footnote{\url{ https://www.leagueofrobotrunners.org/}} sponsored by Amazon Robotics, was organized online in 2023 to examine more challenging LMAPF settings, including:
(1) a large number of agents of up to $10,000$, 
(2) extreme agent density (i.e., the ratio of the number of agents to the number of vertices on the map) of up to $97.7\%$ (as shown in \Cref{fig:front-fig-random}),
(3) a large map of up to $54,320$ vertices (as shown in \Cref{fig:front-fig-sortation}),
(4) limited planning time of 1 second per step, and
(5) a more realistic action model that considers the heading and rotation of the agents.


\begin{table}[!t]
    \centering
    \begin{tabular}{cccc}
\toprule
Map Name & Map Size & Vertices & Map Illustration \\
\midrule
Random & \phantom{1}\phantom{1}32$\times$32 & \phantom{1}\phantom{1} 819  & \Cref{fig:front-fig-random} \\
City  & 256$\times$256 & 47,240  & \Cref{fig:map-city} \\
Game  & 481$\times$530 & 43,151 &  \Cref{fig:disabling_agents_brc292d}\\
Warehouse  & 140$\times$500 & 38,586 & \Cref{fig:wait_heatmap_warehouse} \\
Sortation  & 140$\times$500 & 54,320 &  \Cref{fig:front-fig-sortation} \\
\bottomrule

    \end{tabular}
    \caption{The basic information for each map. Map sizes are given in the format of height $\times$ weight.}
    \label{tab:maps}
\end{table}

The LRR competition leads us to discuss three critical research challenges less explored in the previous LMAPF literature: the limited planning time constraint, the congestion and myopic behaviors in LMAPF algorithms, and the gaps between modeled and real-world LMAPF. Inspired by these challenges, we point out several future directions, such as developing more competitive anytime algorithms, exploiting multi-CPUs and GPUs for parallelism, alleviating congestion and myopia of windowed planning by developing guidance
and traffic rules, incorporating future traffic prediction and real-time search, and determining the optimal agent number. We also advocate for studying more realistic LMAPF models taking into account motion constraints, execution uncertainty, and evolving systems. \citet{salzman2020research} also discuss several LMAPF challenges, such as lifelong learning and distributed planning, which are orthogonal to this work.

We first formulate the LMAPF problem given in the LRR competition in \Cref{section:problem}. Then, in each of \Cref{section:algorithm,section:myopic}, we first discuss a challenge underexplored in existing work, then our solution to the challenge, and finally future directions inspired by the challenge and our solution. In \Cref{section:simtoreal} we discuss an additional challenge regarding the gaps between LMAPF models in the literature and the real world. Our solution ranked first in two out of three tracks, including the overall best track, and second in another one. The code and experiment settings are publicly available.\footnote{https://github.com/DiligentPanda/MAPF-LRR2023.git}

\section{Problem Definition}
\label{section:problem}

\begin{definition}[MAPF]
MAPF takes in a graph $G=(V,E)$, where $V$ are vertices and $E$ are edges, and $n$ agents with their start and goal vertices. At each step, an agent can move to an adjacent vertex or wait at its current vertex. Collisions occur when two agents move to the same vertex or traverse the same edge at the same step. The target is to find collision-free paths for all agents to their goals while minimizing the sum-of-costs, namely the total number of actions taken.
\end{definition}

\begin{definition}[LMAPF]
\label{def:LMAPF}
LMAPF extends MAPF by constantly assigning new goals to agents via a (external) task assigner. The target is to find collision-free paths for all agents while maximizing throughput, namely the average number of goals reached by all agents per step.
\end{definition}

\begin{definition}[MAPF with Rotations (MAPF-R)]
\label{def:MAPFR}
MAPF-R
is a variant of MAPF with different agent states and actions. It models a four-neighbor grid map as graph $G$ and the state of an agent as a location $v \in V$ plus an orientation $o \in \{$East, South, West, North$\}$. At each step, an agent can forward to an adjacent vertex, rotate 90 degrees clockwise or counterclockwise, or wait at its current vertex. 
\end{definition}

The problem given in the LRR competition is a combination of \Cref{def:LMAPF,def:MAPFR}, named LMAPF-R. Further, the task assigner is external to our planner and assigns exactly one new goal to an agent if the agent reaches its current one.

\begin{table}[!t]
    \centering
    \resizebox{\linewidth}{!}{
    \begin{tabular}{c@{\hspace{1pt}}cccccc}
    \toprule
        Instance & Algorithm & $w$ & $h$  & GG & DA  & Score \\
    \midrule
        Random 100 & WPPL & 20 & 1  & No & No & 0.994 \\
        Random 200 & WPPL & 15 & 1 & Manual & No & 0.975\\
        Random 400 & WPPL & 15 & 1 & Manual & No & 0.992  \\
        Random 600 & WPPL & 15 & 1 & Manual & Yes & 1.000 \\
        Random 800 & PIBT & 1 & 1 & GGO & Yes & 0.806 \\
        City 1000 & WPPL & 20 & 1 & No  & No  & 0.996 \\
        City 3000 & WPPL & 20 & 1 & Manual & No  & 1.000\\
        Game 4000 & WPPL & 15 & 3 & Manual & Yes  & 1.000\\
        Warehouse 8000 & WPPL & 10 & 3 & Manual & No & 0.996 \\
        Sortation 10000 & WPPL & 10 & 3 & Manual & No & 0.997\\
    \bottomrule
    \end{tabular}
    }
    \caption{The best hyperparameters for each instance and the best scores of similar-setting online instances. Each instance is named as its map with its number of agents. An algorithm replans length-$w$ paths every $h$ step. GG and DA refer to Guidance Graph and Disabling Agents respectively. Details are explained in \Cref{section:algorithm,section:myopic}. }
    \label{tab:hyperparameter}
\end{table}

\paragraph{Competition and Experiment Setup}
 All results are evaluated on a server with 32 AMD EPYC 7R13 Processor vCPUs and 128G Memory in the LRR competition. For simplicity, in this paper, we evaluate all the results on a local machine with 32 Intel(R) Xeon(R) Platinum 8350C vCPUs and 84G Memory. Since the online data was not publicly available when this paper was written,\footnote{The online data is now available at https://github.com/MAPF-Competition/Benchmark-Archive.git} we conducted our experiments with offline data that share similar settings to the online data. The map of each instance is known before simulation and given 30 minutes for preprocessing. We list all maps in \Cref{tab:maps} and problem instances in \Cref{tab:guidance}. The best hyperparameters are shown in \Cref{tab:hyperparameter}. 

The competition ranks teams with 10 hidden instances in three tracks: Overall Best, Fast Mover, and Line Honours. For each instance, a normalized score between 0 and 1, defined as the throughput of a solution over the best throughput among all solutions submitted by all participants, is calculated. The Overall Best track ranks teams by the total score of all instances in one submission but allows the planner to exceed the 1-second limit. However, the system will lose the chance of moving for several steps until the current planning is finished. The Fast Mover track further restricts all actions to be planned within the time limit. The Line Honours track ranks teams by the total number of best solutions to individual instances but allows these best solutions to come from different submissions. Our final submission has close-to-1 scores on 9 out of 10 instances (\Cref{tab:hyperparameter}) with a total score of 9.755, compared to 9.502 of the second-place team. Our solution won the Overall Best and Fast Mover tracks and ranked second in the Line Honours track. According to what we know, the second-place and third-place teams mainly used independent A* and CBS-based algorithms \cite{Sharon2015cbs}, which are very different from ours.

\begin{figure*}[!t]
    \centering
    \includegraphics[width=0.8\linewidth]{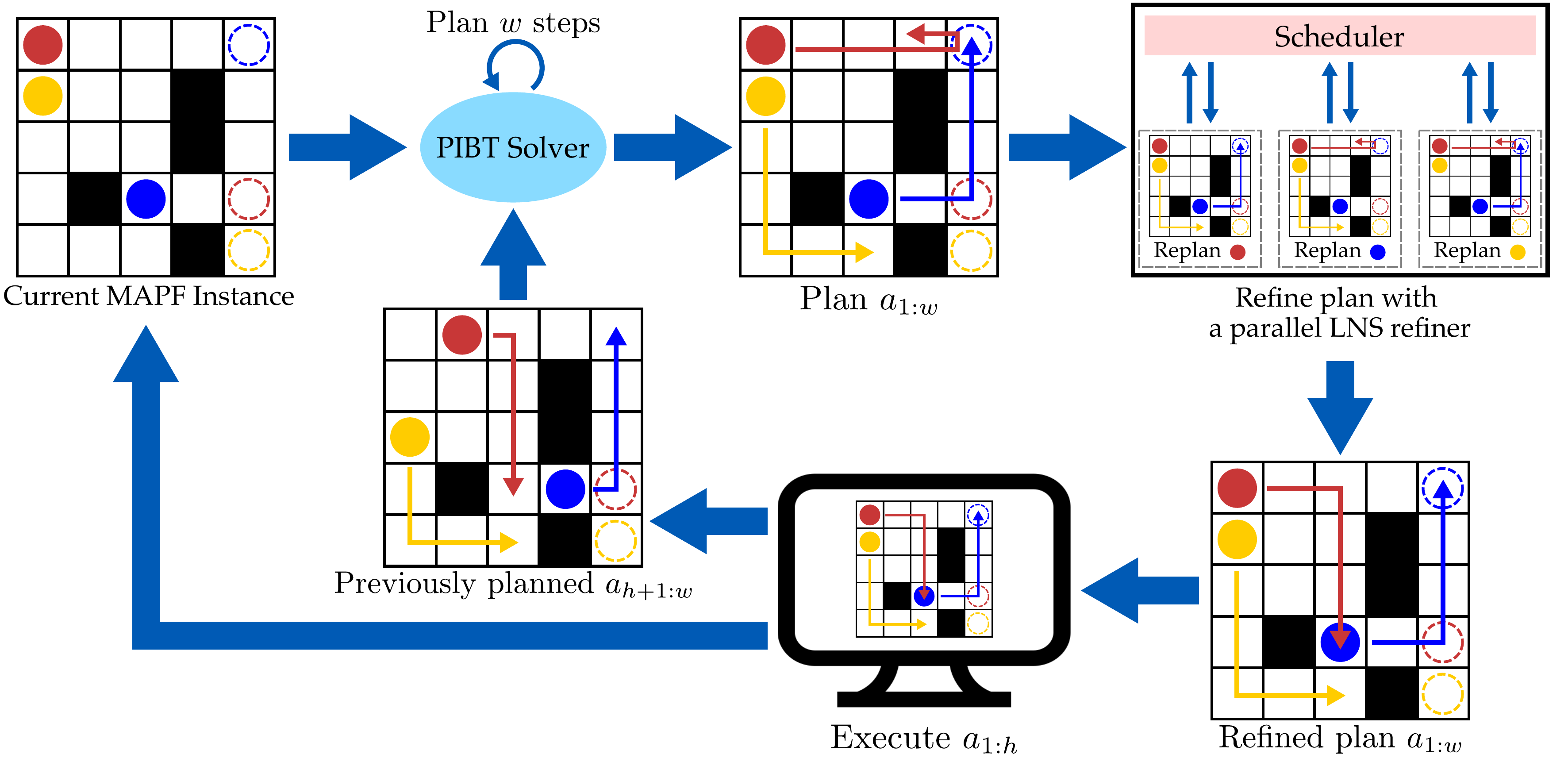}
    \caption{Windowed Parallel PIBT-LNS (WPPL). For each windowed MAPF instance, we run PIBT $w$ steps to get an initial plan $a_{1:w}$ and use Parallel MAPF-LNS to refine it. We then execute the first $h$ actions of the refined plan $a_{1:h}$, updating the MAPF instance and reusing the rest of actions $a_{h+1:w}$ in the next iteration of PIBT. 
    }
    \label{fig:framework}
\end{figure*}


\section{Challenge 1: Limited Planning Time}
\label{section:algorithm}

In this section, we discuss the first challenge of \emph{how to obtain high-quality LMAPF solutions within a limited planning time}. Many real-world applications have such a requirement. For example, smart manufacturing and automated warehouses require efficient execution while autonomous driving and video games need real-time response. The LRR competition poses this challenge by giving only 1 second to plan actions for all agents at each step. This time limit is significantly shorter than the limits used in most existing work. We first analyze the existing work and then present our algorithm choices in the competition. Lastly, we discuss the related future directions.


\subsection{The Challenge and Related Works}



The challenging online LMAPF instance is generally decomposed into a sequence of MAPF instances solved by MAPF solvers. However, repeatedly replanning paths for all agents is extremely time-consuming. Several approaches have been proposed to reduce the computational effort and increase the scalarbility. For example, \citet{WanICARCV18} propose an incremental variant of CBS to reuse the previous search effort. \citet{SvancaraAAAI19} propose to only replan for the agents who receive new goals and the agents whose paths are affected by the former agents. However, the scalability of these methods is still very limited as indicated in the state-of-art work~\cite{Li2021RHCR}, which proposes the RHCR framework. RHCR largely reduces computation time by adopting the idea of windowed planning which does not reason about conflicts beyond a fixed window. However, only an efficient framework is not enough. Indeed, when applying RHCR with PBS~\cite{MaAAAI19} and ECBS~\cite{BarrerSoCS14}, we can solve instances at most with hundreds of agents within the 1-second limit. Therefore, we need a faster MAPF solver that still provides good solutions.


There are various MAPF algorithms, most of which can be categorized as follows. \textbf{Optimal algorithms}, such as CBS~\cite{Sharon2015cbs} and BCP~\cite{lam2020new}, can return optimal solutions but have an exponential increase in runtime.
\textbf{Bounded-suboptimal algorithms}, such as
ECBS~\cite{BarrerSoCS14} and 
EECBS~\cite{li2021eecbs}, 
guarantee solutions within a given bound and scale to hundreds of agents, but fail to tackle more challenging problems. \textbf{Prioritized algorithms}~\cite{Erdmann87} and \textbf{rule-based algorithms}~\cite{okumura2019priority,wang2020walk} run much faster, but are incomplete and lack guarantee on their solution quality. These algorithms fall on the two extremes of a spectrum, either provide high-quality solutions but do not scale to challenging scenarios (such as 10k agents, 97\% density, etc.) (most MAPF algorithms), or scale to them but with poor solution quality (such as PIBT). 

In the competition, we found two other types of algorithms, namely anytime and parallel algorithms, very useful when facing a limited planning time. However, both categories are underexplored compared to the previous ones. 

\textbf{Anytime algorithms} offer a better trade-off between solution quality and runtime due to the anytime behavior, which refers to first generating an initial solution and then improving it over time. However, most existing works either cannot scale to a large number of agents~\cite{vedder2021x} 
or have very weak anytime behaviors ~\cite{lam2020new,okumura2023lacam}.
The state-of-the-art anytime algorithm for MAPF is MAPF-LNS~\cite{li2021anytime}. Starting with an initial solution, MAPF-LNS continuously selects a group of agents and replans their paths. 

\textbf{Parallel algorithms} are designed to exploit multi-core CPUs or GPUs. \citet{lee2021parallel} parallelize CBS by solving subproblems in parallel, then merging the search trees of subproblems.
\citet{li2021scalable} run multiple MAPF-LNS on independent threads and select the best solution among them. \citet{leet2022shard} divide the map into multiple regions and solve MAPF in each region in parallel.

\subsection{Our Solution: WPPL}

We present Windowed Parallel PIBT-LNS (WPPL) as our solution to address the challenging LMAPF instances in the LRR competition. We give an overview of WPPL in \Cref{fig:framework} and introduce the design choices of each component below.


\noindent \textbf{Windowed Planning.} With large agent numbers and map sizes, even planning the full shortest paths individually for all agents is computationally challenging within 1 second. Therefore, we apply the idea of a planning window in 
RHCR \cite{Li2021RHCR} to replan all agents every $h$ steps with a planning window of length $w$ ($w \geq h$). Specifically, at step $0$, we use the initial $1$ second to plan a path of length $w$ for each agent. Later, we always use the $h$ seconds from step $kh+1$ to $(k+1)h$, $k\in \mathbb{N}$, to plan the path of length $w$ starting from the state at step $(k+1)h$.



\begin{figure*}[!t]
    \centering
    \includegraphics[width=0.8\textwidth]{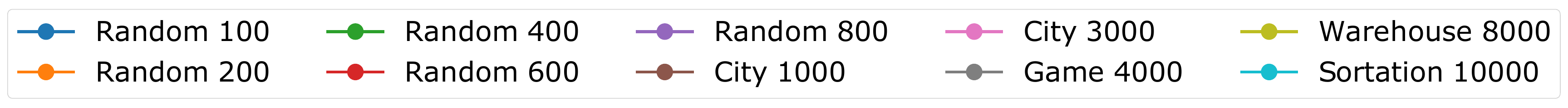}\\
    \begin{subfigure}[b]{0.18\textwidth}
      \centering
      \includegraphics[width=1\textwidth]{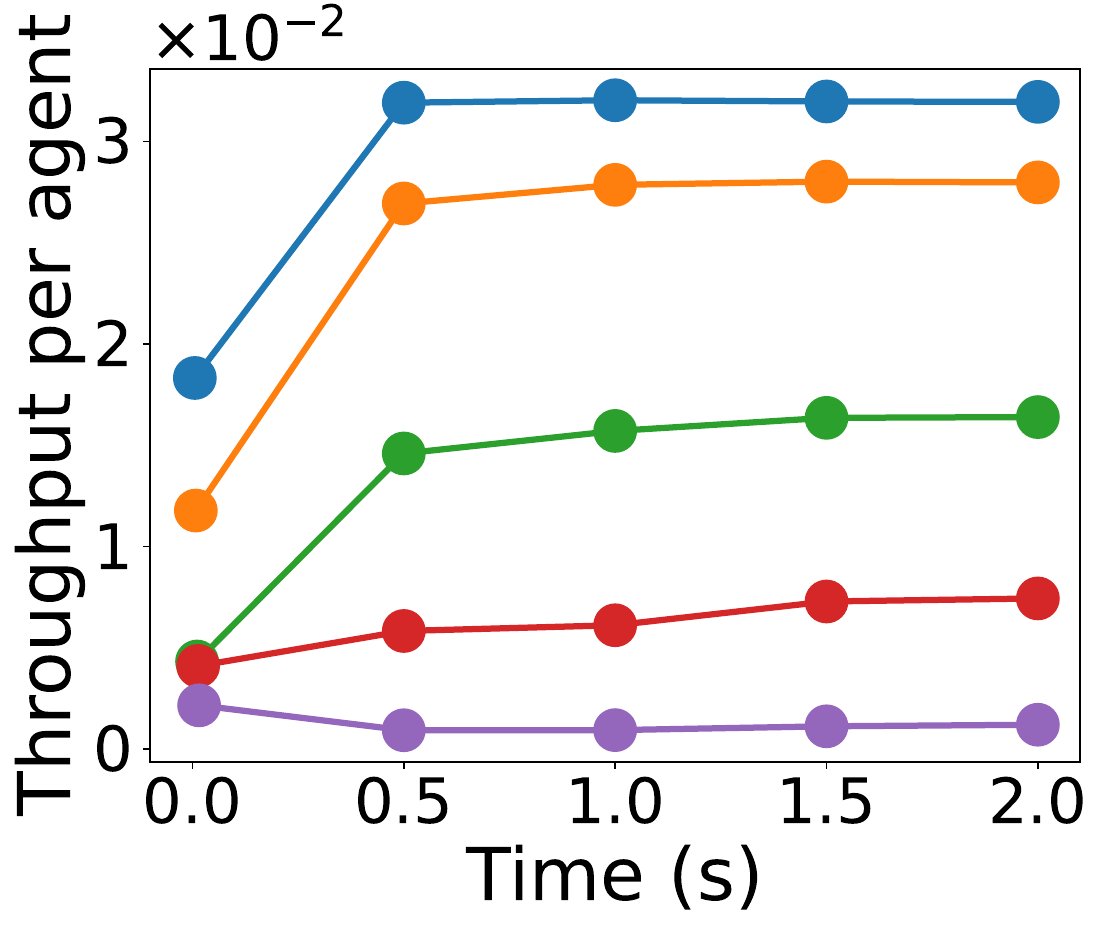}
      \caption{Random}
    \end{subfigure}%
    \hfill
    \begin{subfigure}[b]{0.19\textwidth}
      \centering
      \includegraphics[width=1\textwidth]{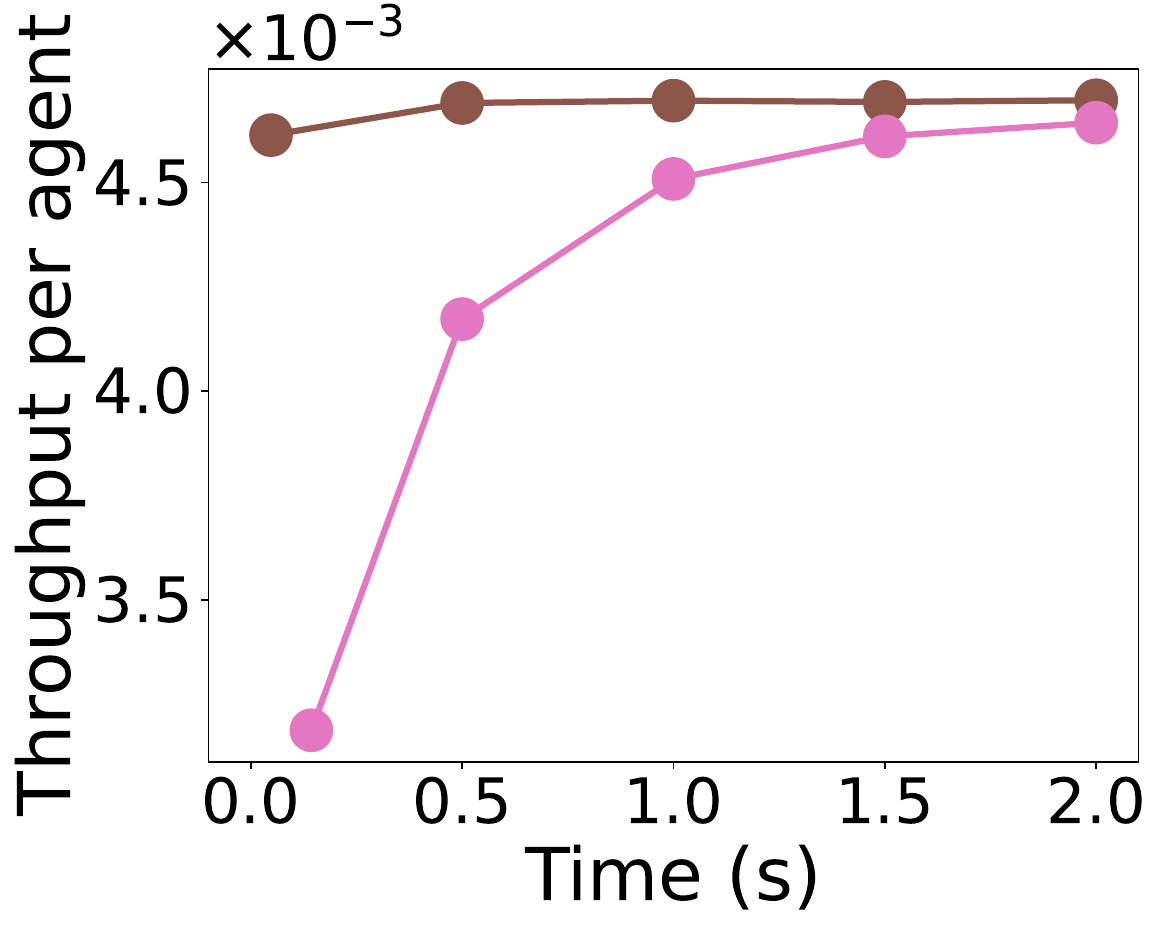}
      \caption{City}
    \end{subfigure}%
    \hfill
    \begin{subfigure}[b]{0.1925\textwidth}
      \centering
      \includegraphics[width=1\textwidth]{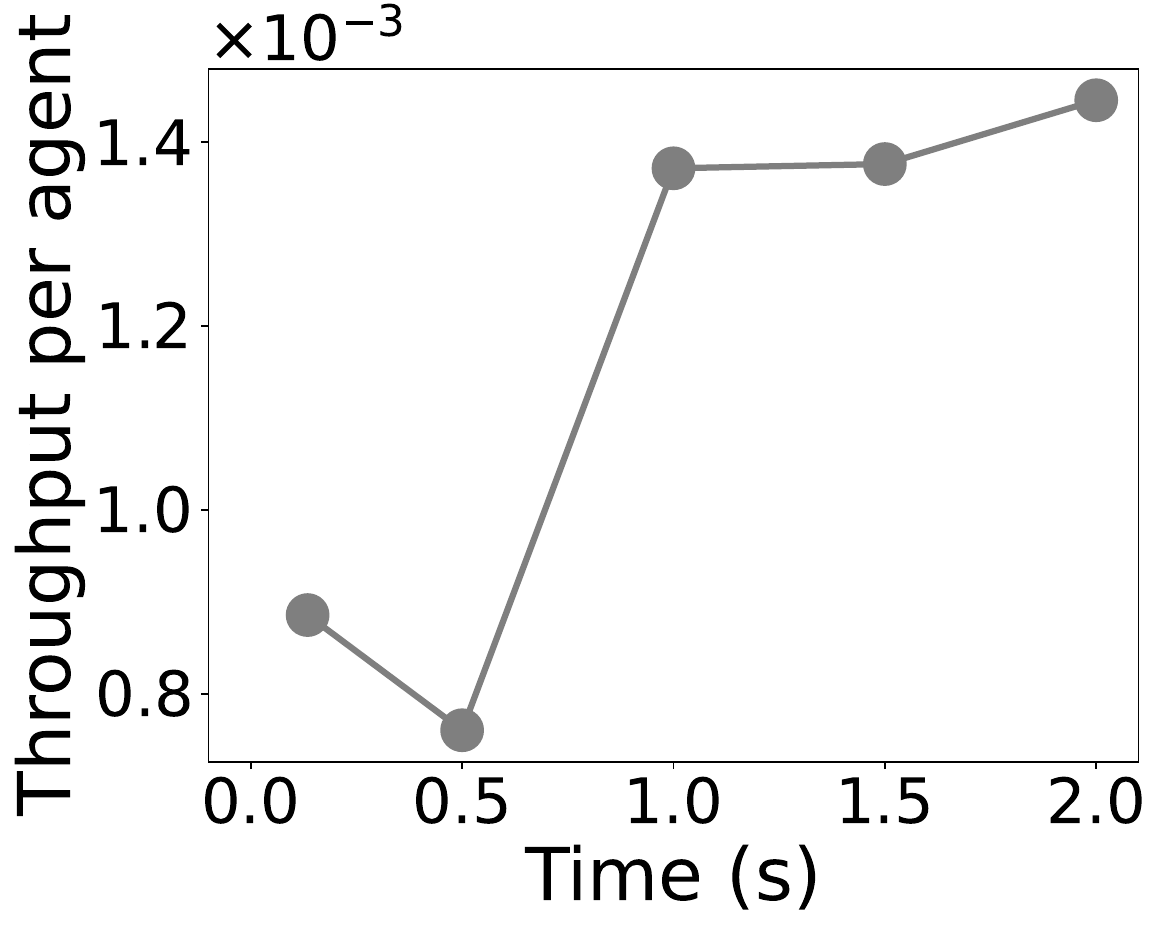}
      \caption{Game}
    \end{subfigure}%
    \hfill
    \begin{subfigure}[b]{0.205\textwidth}
      \centering
      \includegraphics[width=1\textwidth]{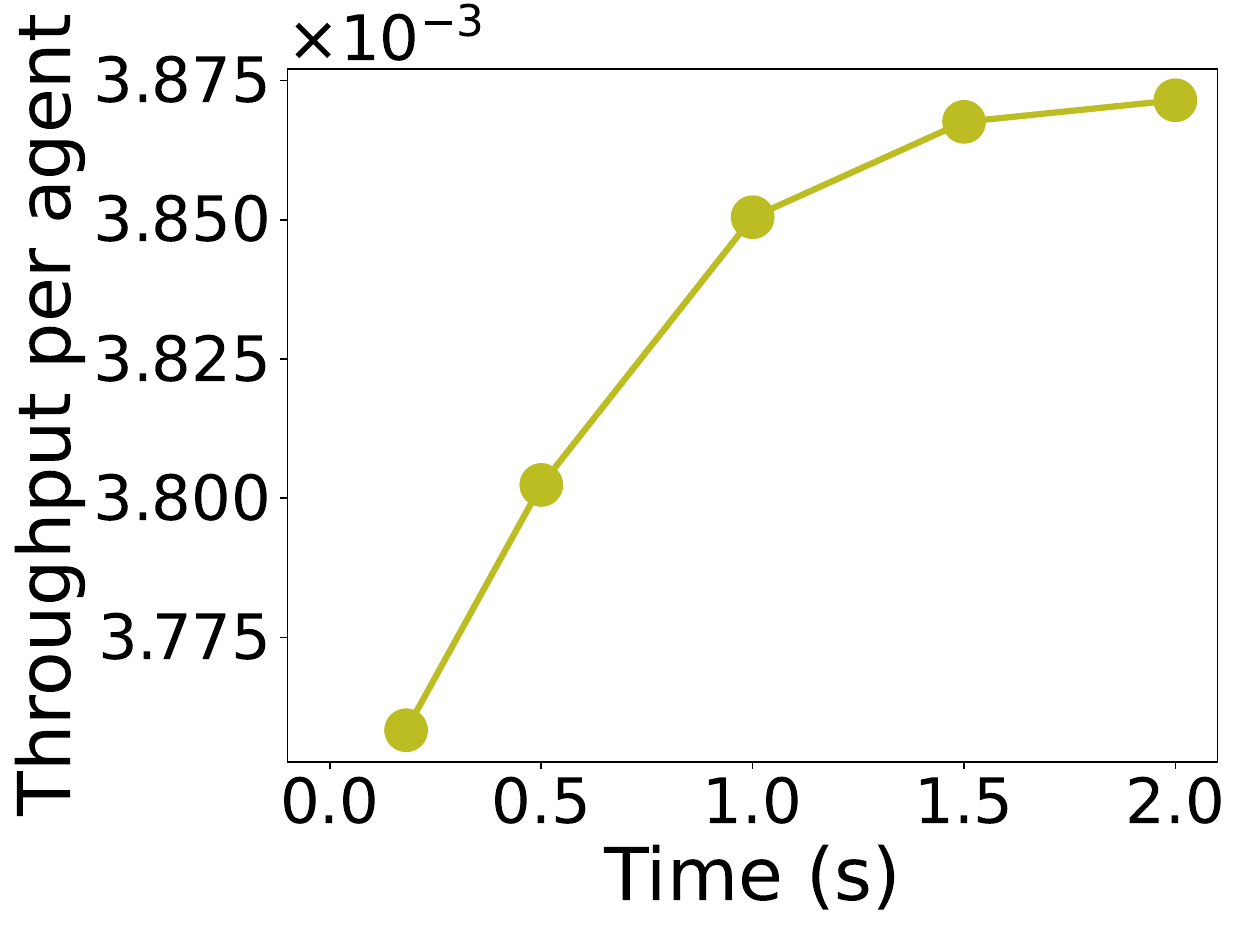}
      \caption{Warehouse}
    \end{subfigure}%
    \hfill
    \begin{subfigure}[b]{0.1925\textwidth}
      \centering
      \includegraphics[width=1\textwidth]{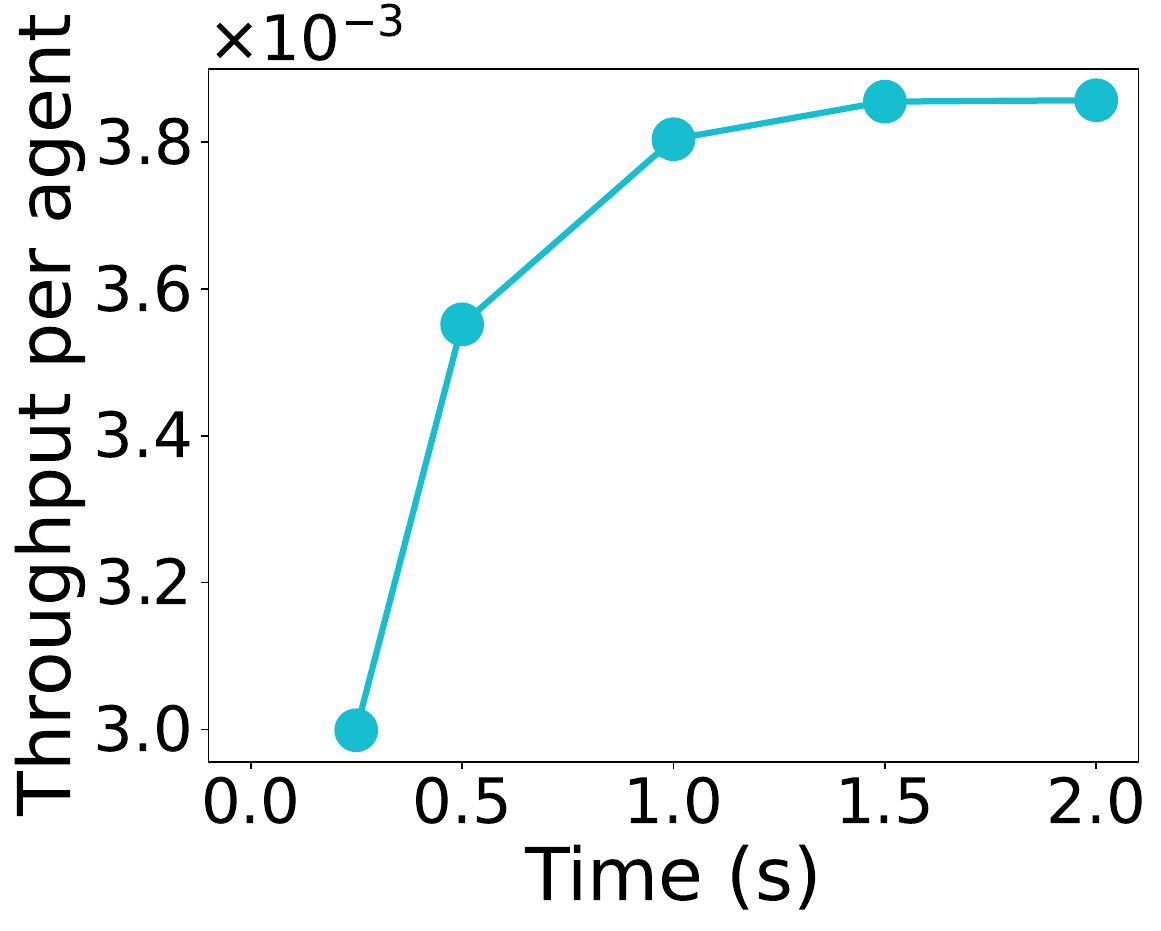}
      \caption{Sortation}
    \end{subfigure}%
    \hfill
    \caption{The anytime behavior of WPPL. 
    For better illustration, we apply guidance graphs (introduced later in \Cref{section:myopic}), use only a single thread, and normalize throughput by the number of agents in this experiment.
    }
    \label{fig:anytime}
\end{figure*}

\noindent \textbf{PIBT-LNS.} Our windowed MAPF solver is PIBT-LNS, which obtains an initial solution using PIBT~\cite{okumura2019priority}
and then applies MAPF-LNS~\cite{li2021anytime} to improve it.
We select PIBT because it is an extremely fast algorithm and the only algorithm we know that can handle very large and dense problem instances. 
In our experiments, PIBT returns a solution within 250 ms in the large-scale Sortation 10000 instance and within 20 ms in the dense Random 800 instance, leaving ample remaining runtime. 
Therefore, we use the anytime algorithm MAPF-LNS to improve the solution and fully exploit the given planning time.
To illustrate the effectiveness of MAPF-LNS, we run experiments with different planning time limits in \Cref{fig:anytime} and observe progressive improvements in throughput for all instances except for Random 800 and Game 4000 at 0.5 seconds. We will discuss this abnormal behavior in \Cref{section:myopic}. 



\begin{figure*}[!t]
    \centering
    \begin{subfigure}[b]{0.18\textwidth}
      \centering
      \includegraphics[width=1\textwidth]{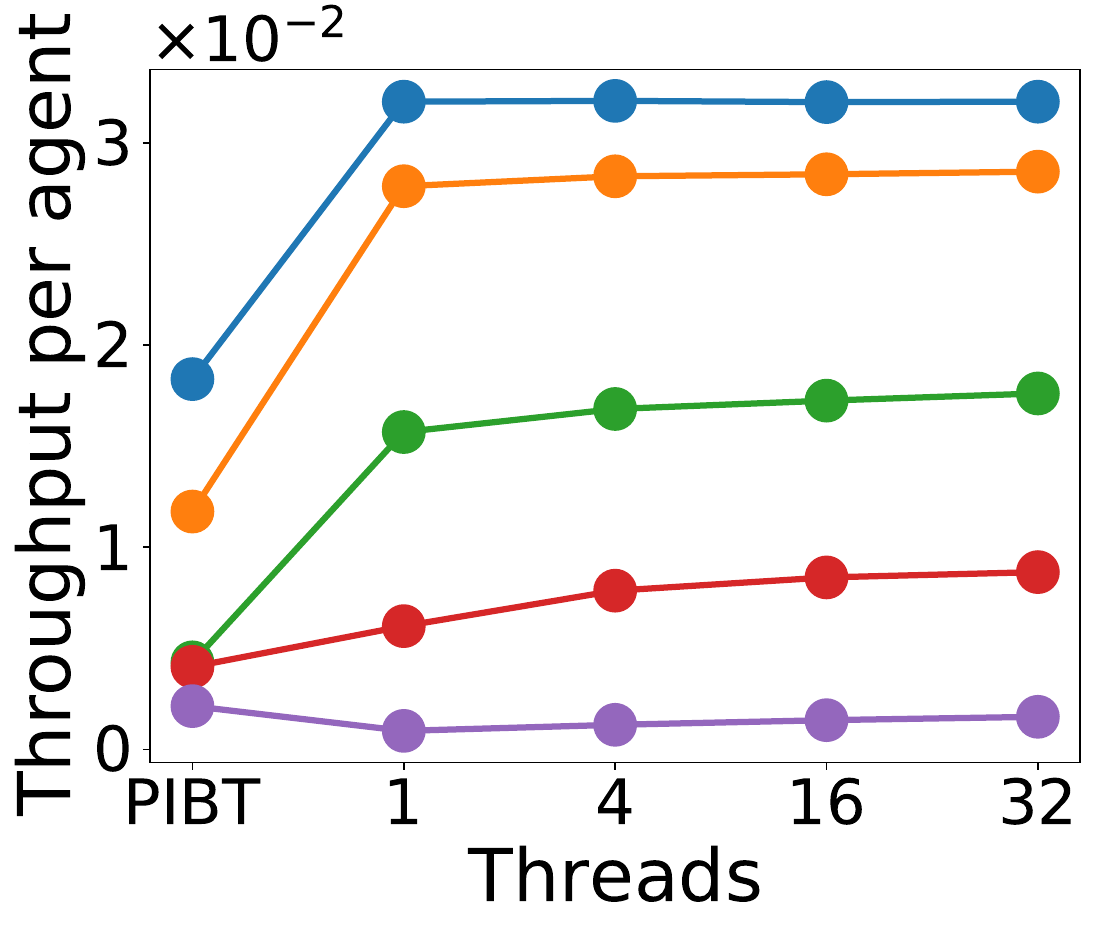}
      \caption{Random}
    \end{subfigure}%
    \hfill
    \begin{subfigure}[b]{0.19\textwidth}
      \centering
      \includegraphics[width=1\textwidth]{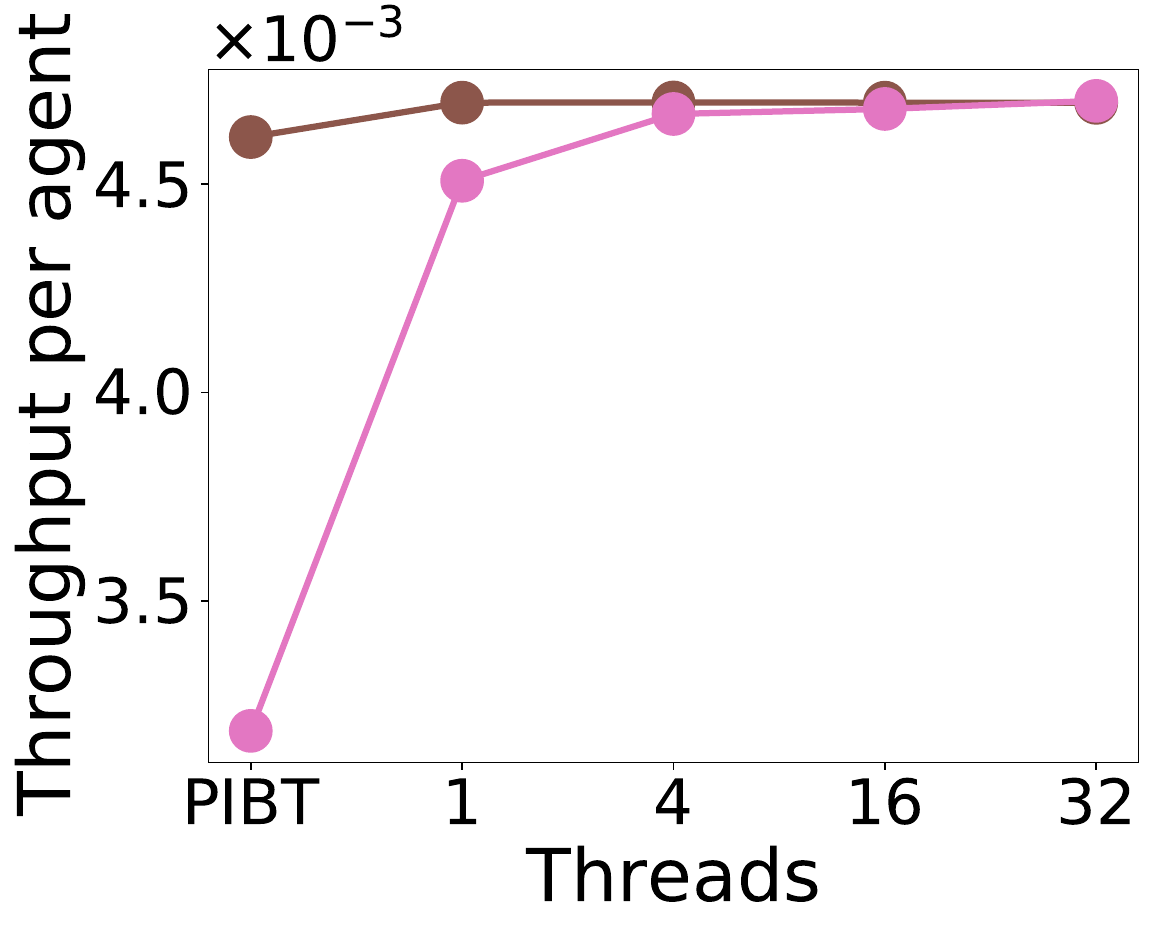}
      \caption{City}
    \end{subfigure}%
    \hfill
    \begin{subfigure}[b]{0.19\textwidth}
      \centering
      \includegraphics[width=1\textwidth]{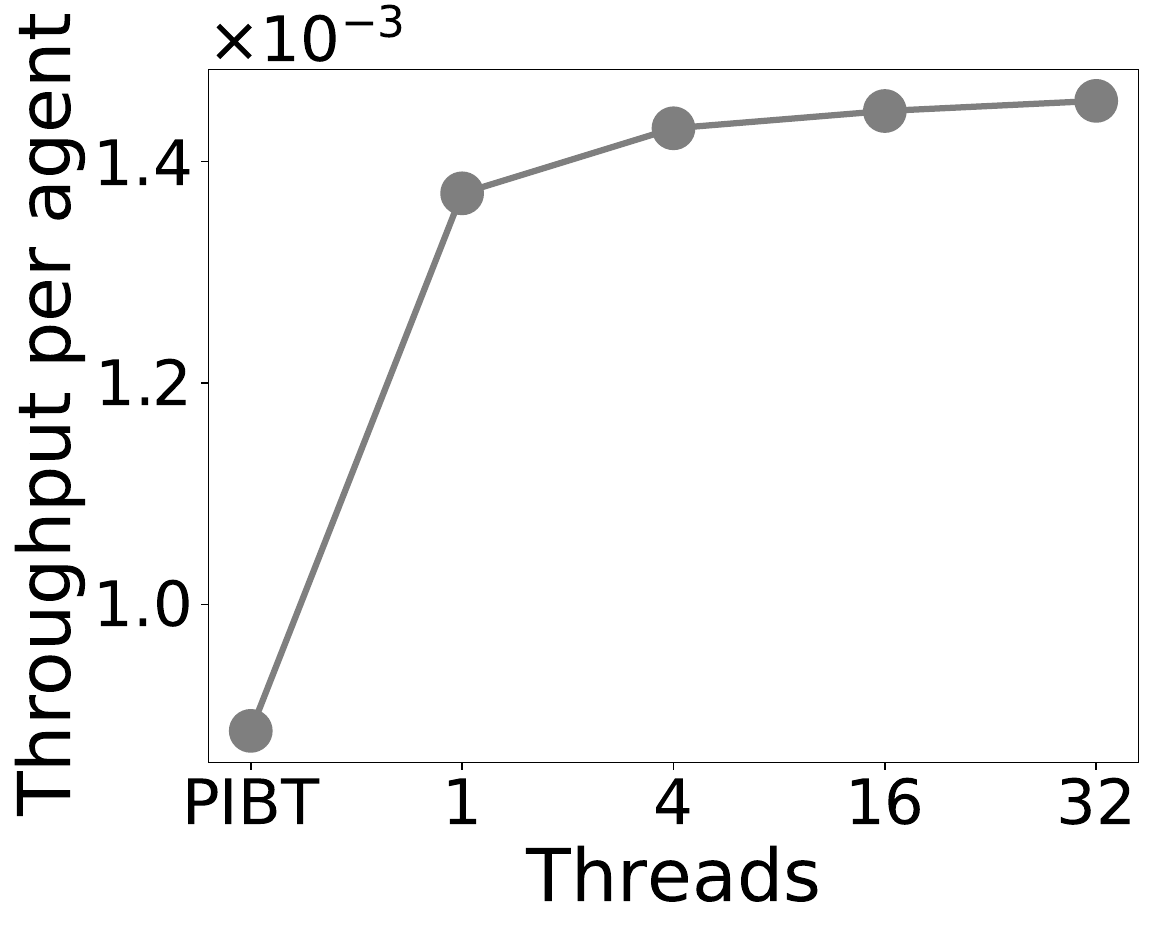}
      \caption{Game}
    \end{subfigure}%
    \hfill
    \begin{subfigure}[b]{0.195\textwidth}
      \centering
      \includegraphics[width=1\textwidth]{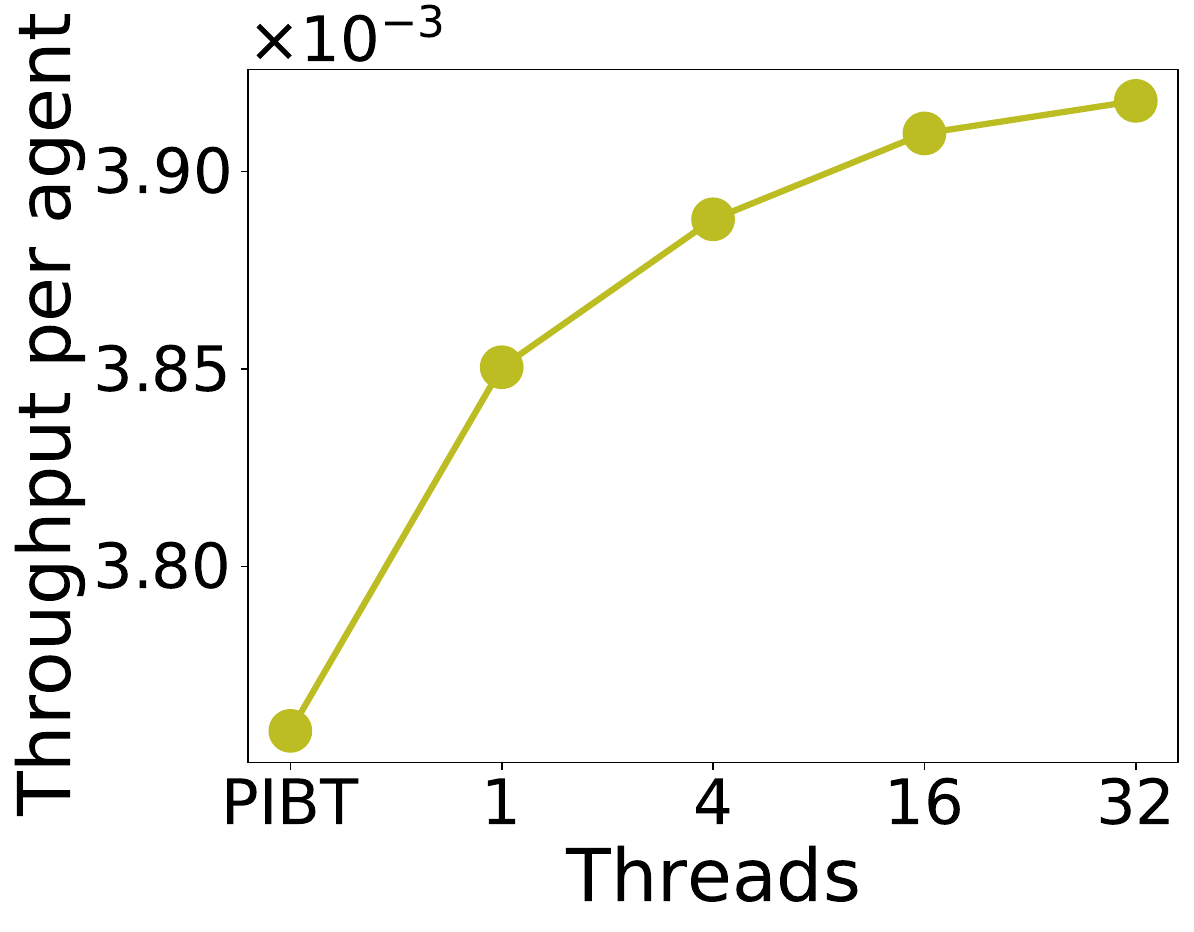}
      \caption{Warehouse}
    \end{subfigure}%
    \hfill
    \begin{subfigure}[b]{0.195\textwidth}
      \centering
      \includegraphics[width=1\textwidth]{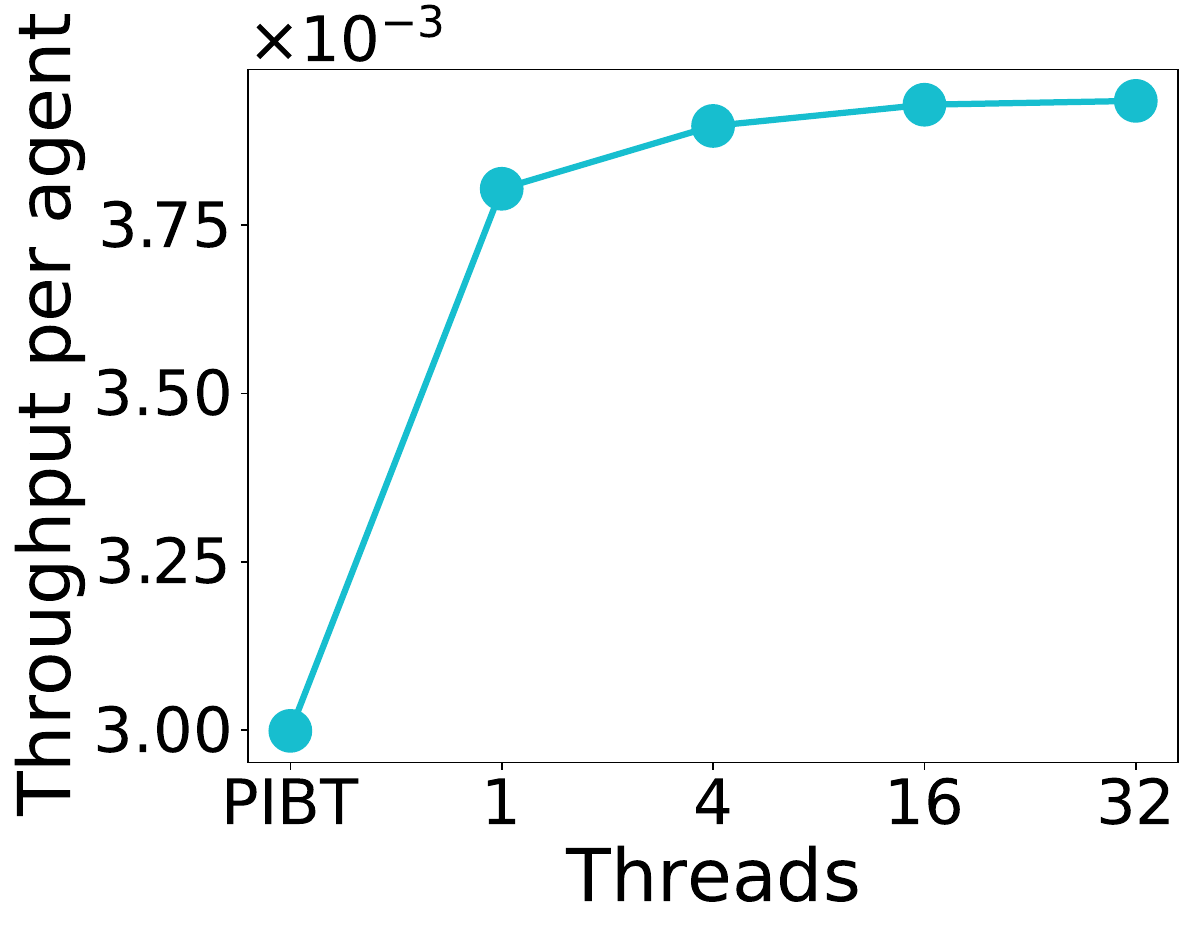}
      \caption{Sortation}
    \end{subfigure}%
    \hfill
    \caption{The effect of Parallelizing LNS. 
    For better illustration, we apply guidance graphs (introduced later in \Cref{section:myopic}) and normalize throughput by the number of agents  in this experiment. This figure uses the same legend as \Cref{fig:anytime}.
    }
    \label{fig:parallel}
\end{figure*}

\noindent \textbf{Parallel PIBT-LNS.} We further parallelize PIBT-LNS to exploit multi-cores on the evaluation server. The parallelization strategy that we applied shares a great similarity with a concurrent work, DROP-LNS~\cite{chan2024anytime}. We asynchronously select groups of agents and replan paths for each group in parallel, but sequentially check each new sub-plan and update better ones to the global plan. We show the effectiveness of parallelization in \Cref{fig:parallel}.

\noindent \textbf{Reuse Previous Search Effort.} Since we repeatedly solve overlapping windowed MAPF instances, it is natural to think of reusing search efforts from the previous planning. 
\citet{WanICARCV18} and \citet{madar2022leveraging} successfully apply this idea to reduce the search time of CBS and PBS respectively in the lifelong setting. However, in our case, we hope that a partial plan refined by the previous search could provide hints for PIBT to obtain a better-quality initial solution. In our ablation study, reusing the past $w-h$ planned steps improves the throughput of Random 400, Random 600, and Random 800 by 4.2\%, 29.9\%, and 4.9\%, respectively. For other maps, the improvement is below 2\%. We conjecture it is because PIBT tends to find bad initial solutions in unstructured maps and crowded scenarios.

\subsection{Future Directions}

When designing WPPL to solve the LMAPF instances in the LRR competition, we discover several directions worth further study.


\noindent \textbf{Better Rule-based Algorithms.}
We use PIBT to find an initial solution to a given MAPF instance. As a rule-based algorithm, PIBT has a faster speed but significantly lower solution quality than optimal and bounded-suboptimal MAPF algorithms~\cite{li2022lns2}. Thus, one direction is to develop rule-based algorithms with better solution quality. In addition, Multi-Agent Reinforcement Learning (MARL) is an emerging and promising alternative to rule-based algorithms. PRIMAL$_2$~\cite{DamaniRAL21}, a state-of-the-art MARL framework for LMAPF, has demonstrated a faster runtime but lower solution quality than RHCR. Therefore, we can also focus on improving the solution quality and addressing the incompleteness issue of MARL methods.


\noindent \textbf{Better Anytime Algorithms.}
Anytime algorithms are desirable in practice because they can make full use of the planning time budget. One straightforward idea is to improve MAPF-LNS, for example, by developing better ways to select groups of agents~\cite{huang2022anytime}. Another possibility is to improve the ability of other existing anytime algorithms like anytime ECBS~\cite{ilya2022anytimeECBS} to rapidly improve performance within a short amount of time. Meanwhile, future studies can also work on extending other single-agent anytime algorithms, such as ARA*~\cite{Likhachev2003ARAAA}, to multi-agent scenarios.

\noindent \textbf{Parallelism for MAPF.}
With the rapid development of technologies, multi-CPUs and GPUs become cheaper, faster, and more prevalent. Other fields like deep learning and evolutionary computation have developed efficient software to leverage them. However, the MAPF community has fallen behind in exploiting parallelism due to the natural challenge of parallelizing centralized search algorithms. While some works have explored parallelizing CBS~\cite{lee2021parallel} or LNS~\cite{chan2024anytime} on multi-CPUs,
the problem of efficiently parallelizing state-of-the-art MAPF algorithms stays underexplored. 
Therefore, future work can also study how to parallelize these search-based methods.


\section{Challenge 2: Traffic Congestion and the Myopic Behavior of LMAPF Algorithms}
\label{section:myopic}

In this section, we discuss the second challenge of \emph{how to alleviate traffic congestion and the effect of myopic planning}. We start by discussing the myopic behavior of LMAPF algorithms and analyze its relation to congestion. Then we present our methods, using guidance graphs~\cite{zhang2024guidance} and disabling agents to address these issues. Finally, we discuss the future work.

\subsection{The Challenge and Related Work} \label{sec:myopic-challenge}

\begin{table*}[!t]
    \centering
    \begin{tabular}{cccrrrrrr}
\toprule
\multirow{ 2.5}{*}{Instance} & \multirow{ 2.5}{*}{Total Steps} & \multirow{ 2.5}{*}{Agent Density} & \multicolumn{2}{c}{No GG} & \multicolumn{2}{c}{Manual GG}  & \multicolumn{2}{c}{Auto GG} \\
 \cmidrule{4-9}
  & & & PIBT & WPPL & PIBT & WPPL & PIBT & WPPL \\
\midrule
Random 100 & \phantom{1} 500 & 12.2\%  & 2.21 & \textbf{3.20} & 1.83 & 3.08 & 2.64 &  3.11\\
Random 200 & \phantom{1} 500 & 24.4\%  & 2.55 & 5.69 & 2.35 & \textbf{5.72} & 2.74 &  5.56\\
Random 400 & 1,000           & 48.9\%  & 1.58 & 4.79 & 1.73 & \textbf{7.04} & 1.77 &  4.86\\
Random 600 & 1,000           & 73.4\% & 1.81 & 2.44 & 2.46 & \textbf{5.26} & 2.53 &  2.57\\
Random 800 & 2,000           & 97.7\%  & 1.34 & 1.04 & 1.56 & 1.47 & \textbf{1.72} &  1.23\\
City 1000 & 2,000 & \phantom{1}2.1\%   & 4.39 & \textbf{4.72} & 4.61 & 4.70 & - & -\\
City 3000 & 4,000 & \phantom{1}6.4\%   & 5.95 & 6.13 & 9.57 & \textbf{14.10} & - & -\\
Game 4000 & 5,000 & \phantom{1}9.3\%   & 2.01 & 1.18 & 3.54 & \textbf{5.45} & - & -\\
Warehouse 8000 & 5,000 & 20.7\%        & 10.37 &5.89 & 30.07 & \textbf{31.34} & - & -\\
Sortation 10000 & 5,000 & 18.4\%       & 10.20 & 10.19 & 29.99 & \textbf{39.34} & - & -\\
\bottomrule
    \end{tabular}
    \caption{
    The table's first three columns show each instance's basic information. 
    Agent density is the agent number divided by the vertex number. The rest of the table shows the throughput of PIBT and WPPL on each instance with different guidance graphs (edge weights) and algorithms. No GG, Manual GG, and Auto GG refer to no guidance graph, manually defined graph, and guidance graph found by GGO algorithm with PIBT respectively. The best throughput is marked in bold. The empty values in Auto GG are due to the large computational demand making GGO unable to run on those maps. 
    }
    \label{tab:guidance}
\end{table*}

\begin{figure*}[!t]
    \centering
    \includegraphics[width=1.0\linewidth]{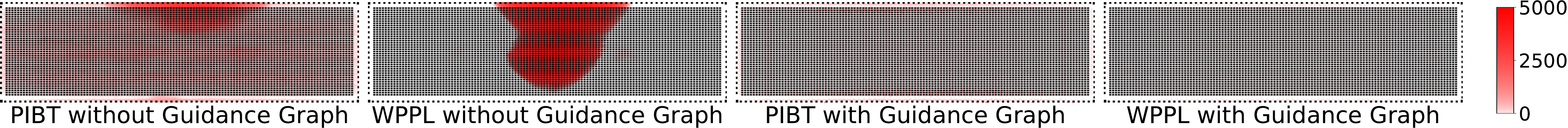}
    \caption{Comparison of with and without guidance graph in the Warehouse 8000 instance. The heatmap shows the wait action usage (the number of steps agents wait in each vertex). Red denotes areas of high congestion.}
    \label{fig:wait_heatmap_warehouse}
\end{figure*}

Congestion is a challenging problem in many lifelong applications, such as traffic and communication systems. Without a proper coordination mechanism, the situation will generally deteriorate over time. We also observed such a phenomenon in the LRR competition. While WPPL obtains competitive throughput on several maps, we find severe congestion on others. The first and second graphs of \Cref{fig:wait_heatmap_warehouse} highlight severe congestion with both PIBT and WPPL in the Warehouse 8000 instance. In addition, while WPPL is expected to refine solutions of PIBT, we observe that WPPL causes more congestion than PIBT and has lower throughput in Random 800, Game 4000, Warehouse 8000, and Sortation 10000, as shown in the no Guidance Graph (GG) column of \Cref{tab:guidance}. In the second graph (WPPL) of \Cref{fig:wait_heatmap_warehouse}, we can also identify more severe congestion in the center of the warehouse, compared to the first graph (PIBT), where the congestion is less severe and more uniform.

We believe that the myopic behavior of LMAPF algorithms is one cause of the congestion. In particular, most LMAPF algorithms, including WPPL, split the LMAPF problem into a sequence of MAPF instances and solve each one with a limited horizon by ignoring future collisions and congestion. Eventually, an optimal solution to the current MAPF instance may form or worsen congestion in the future and may not lead to global optimality.

\begin{figure}[!t]
    \centering
    \includegraphics[width=0.7\linewidth]{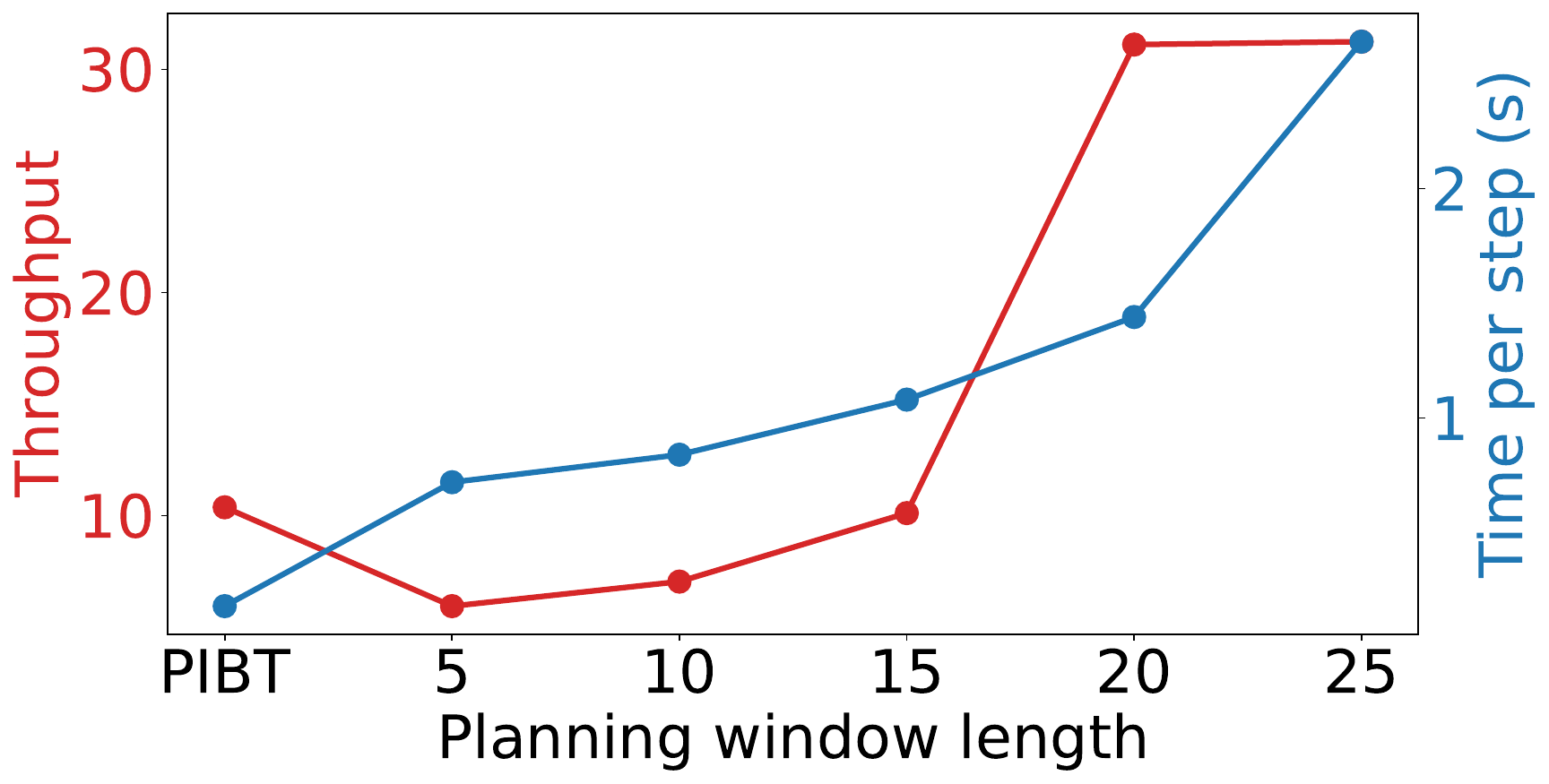}
    \caption{The throughput (red) and the average planning time per step (blue) of WPPL with different window sizes and PIBT in the Warehouse 8000 instance.}
    \label{fig:window_sortation}
\end{figure}

In the case of WPPL, we solve a sequence of windowed MAPF instances by minimizing the approximated sum-of-costs objective, computed as the sum of the actual costs in the window and the approximated costs beyond the window (estimated by the costs of agents' individual shortest paths to their current goals). The approximated objective could be inaccurate if the actual future path costs are much larger than the shortest path approximations. As a result, even though WPPL improves its approximated sum-of-costs objective in the earlier MAPF instances, the movement of agents might cause more severe congestion, and the overall throughput is not guaranteed to improve. Meanwhile, if there is more congestion, the approximated objective could potentially become more inaccurate, making the myopic effect more severe. Namely, the effect of myopia and the congestion may mutually enhance each other. To demonstrate the effect of windowed planning in WPPL, we measure throughput in the Sortation 10,000 instance by running WPPL with various window sizes but fixed LNS refinement iterations to roughly 25,000 in \Cref{fig:window_sortation}. We observe that both the throughput and runtime of WPPL increase with a larger window size $w$. Yet, WPPL does not outperform PIBT until $w > 15$. 

Some methods have been proposed to estimate the future cost better. \citet{Yu2023} apply a deep learning method to predict congestion delays at all vertices for multiple future time windows. \citet{ChenAAAI24} adopt an online estimation approach by hand-crafting heuristics of vertex and contraflow congestion, and iteratively replanning paths.





\subsection{Designing Guidance and Disabling Agents}

As shown in \Cref{sec:myopic-challenge}, while increasing the window size of WPPL can alleviate its myopic behavior, solving the MAPF instance in a larger window takes more time. 
Therefore, instead of trying to design less myopic LMAPF algorithms, we intend to directly alleviate congestion by applying the guidance graph~\cite{zhang2024guidance} and disabling agents. Intuitively, these methods do not change the myopic nature of WPPL. However, they make the myopic behavior of WPPL less harmful to overall throughput.


\noindent \textbf{Design Guidance.} 
We use both manually designed and automatically optimized guidance graphs that are essentially edge weights encoding the costs of moving from one vertex to another (including staying at the original vertex). To manually design the guidance graphs, we start with the crisscross highway~\cite{lironPhDthesis} and fine-tune the edge weights. We slightly increase the weights in congested regions and decrease the weights in idle regions by observing the wait action usage map (e.g., \Cref{fig:wait_heatmap_warehouse}) obtained from the simulation. If throughput increases, we keep the new weight; otherwise, we try different weight modifications. However, such manual fine-tuning is very challenging for the Random map with irregular layouts and almost impossible for the instance Random 800 with an extremely high agent density, where every location seems very congested. Thus, we additionally use the guidance graph optimization (GGO) methods~\cite{zhang2024guidance} to optimize the edge weights automatically. Due to the large computational demand of GGO methods, we cannot run GGO directly with WPPL or on large maps. Therefore, we only experiment GGO with PIBT for instances on the Random map and evaluate the same optimized guidance graph with WPPL in \Cref{tab:guidance}.


\begin{figure}[!t]
    \centering
    \includegraphics[width=0.64\linewidth]{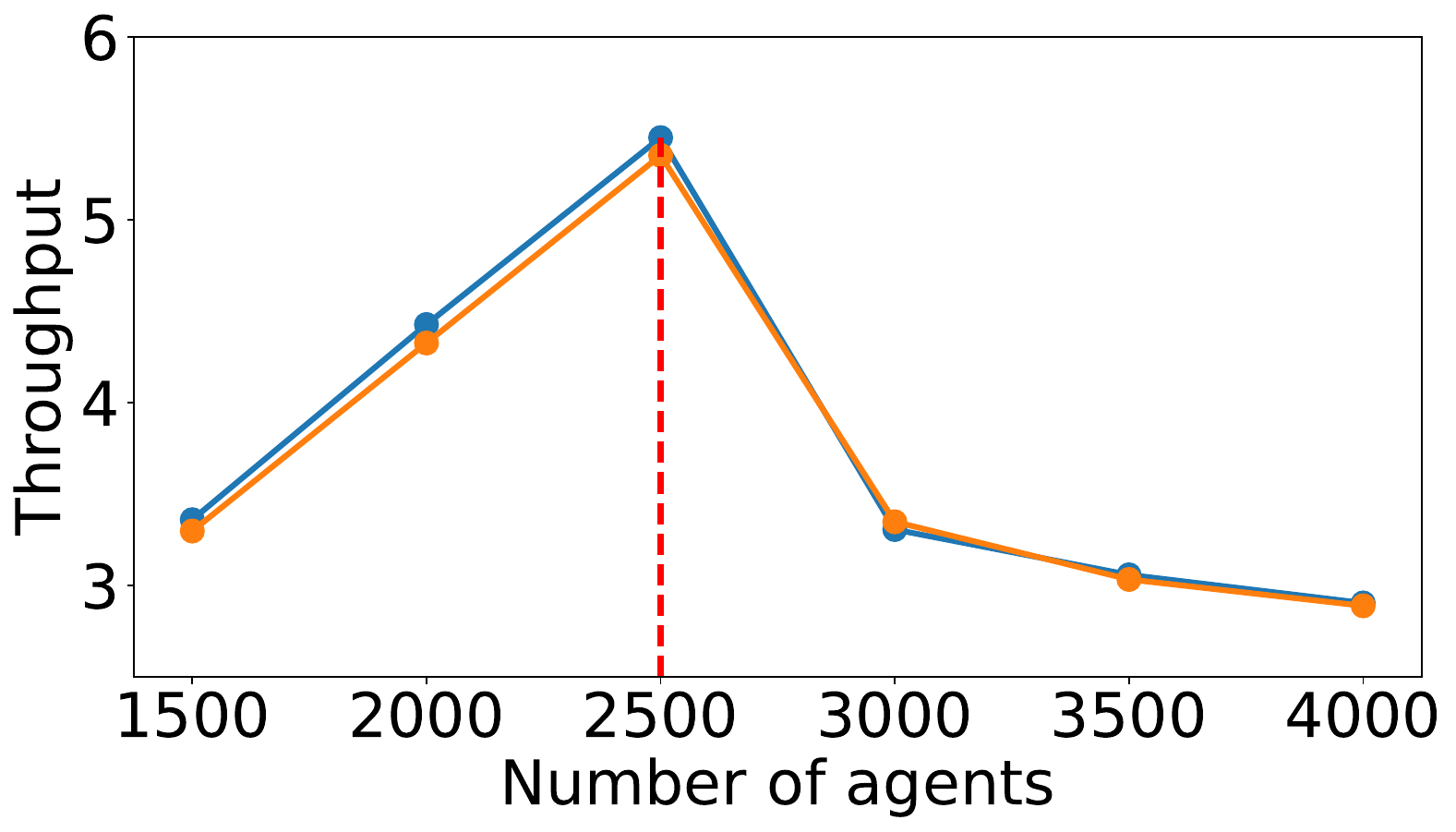}
    \caption{The blue line shows the throughput on the Game 4000 instance with different numbers of agents not disabled (out of 4000 agents). The orange line shows the throughput on the Game map with different total numbers of agents (no agent is disabled). The red dashed line indicates the maximum throughput and the corresponding number of agents.}
    \label{fig:num_agents_game}
\end{figure}

\begin{figure}[!t]
    \centering
    \includegraphics[width=0.9\linewidth]{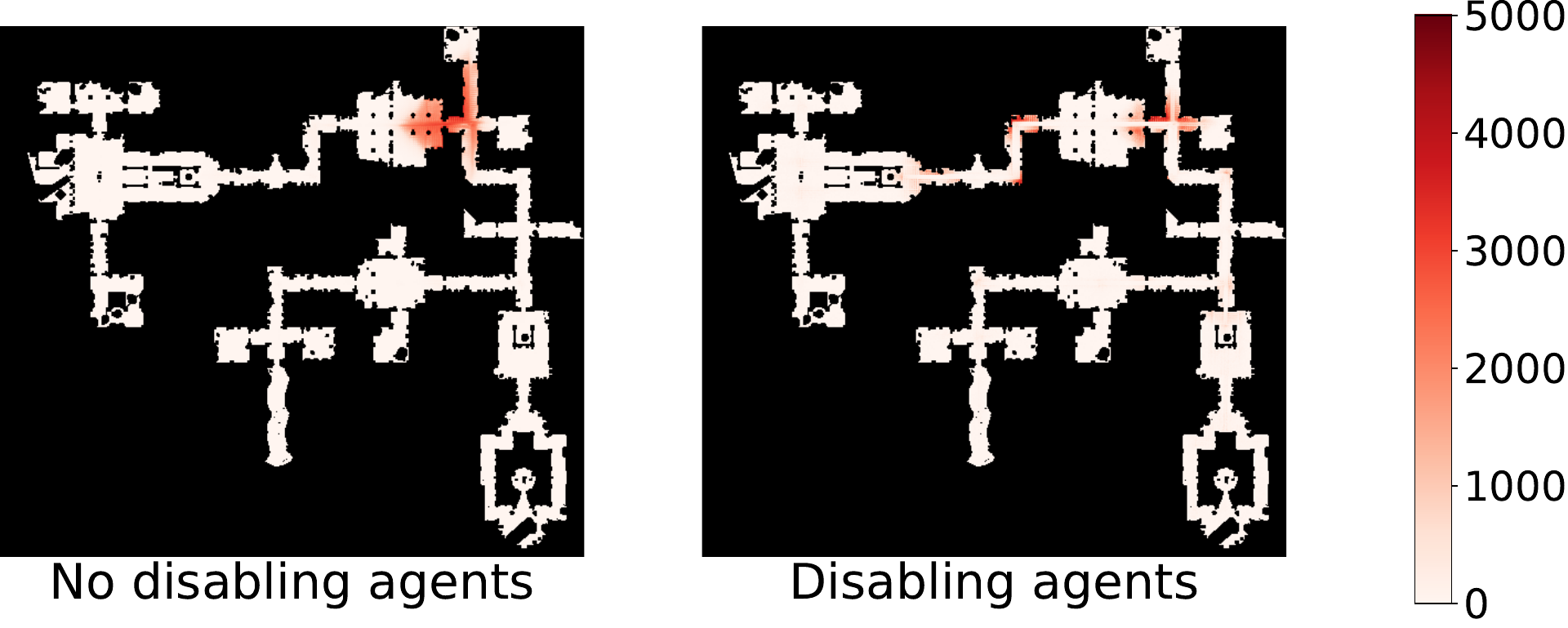}
    \caption{Comparison of with and without disabling agents for the Game 4000 instance. The heatmap shows the wait action usage (the number of times agents wait in each vertex).}
    \label{fig:disabling_agents_brc292d}
\end{figure}



\noindent \textbf{Disabling Agents.} 
Even after we add guidance graphs, we still observe congestion in specific maps, for example, 
at deadends (vertices with only one neighbor) and narrow corridors, especially when there are exceedingly many agents in a local region. However, dealing with geometrically difficult structures remains an open problem. We naively use an ad-hoc trick that disables some agents from going to these regions by setting their goals to their current locations and their priorities to the lowest in both PIBT and the Prioritized Planning algorithm used by MAPF-LNS. As a result, when blocking other higher-priority agents that are not disabled, these disabled agents will be pushed away. For Random 600 and 800, we disable agents whose goals are at the deadends. 
For Game 4000, we search for the most suitable agent number by evaluating various numbers of agents on the map. As shown in \Cref{fig:num_agents_game}, we observe the highest throughput with 2,500 agents, so we randomly disable 1,500 agents. In addition, the coincidence of the blue and orange lines indicates that these disabled agents did not block others due to our changes to their goals and priorities. In the ablation study, the algorithm without disabling agents experiences 15.7\%, 51.8\%, and 47.0\% throughput drops on Random 600, Random 800, and Game 4000, respectively. \Cref{fig:disabling_agents_brc292d} compares the wait action usage before and after disabling agents for the Game 4000 instance, indicating the alleviation of congestion.

\subsection{Future Directions}






Unfortunately, it is impossible to eliminate the myopic behavior of LMAPF algorithms, since it is impractical to know all future tasks. But we can still alleviate it, for example, by traffic predictions, or make its effect less harmful, for example, by reducing congestion. We list several potential solutions in the following.

\noindent \textbf{Better Guidance Graphs and System Designs.}
Since it is challenging or even impractical to manually design guidance for LMAPF instances with arbitrary map size and number of agents, automatic guidance design is appealing. However, GGO still suffers from large computation currently. Therefore, future work can focus on scaling GGO methods to larger-scale problems. In addition, the guidance graphs are optimized offline in GGO, while previous works have explored updating the guidance online during the execution of the LMAPF algorithm~\cite{ChenAAAI24}. Future works can incorporate the online update mechanism into guidance graphs. Furthermore, future work can also focus more broadly on better system design, for example, exploring layout optimization to improve throughput~\cite{zhangLayout23,ZhangNCA2023}. Another idea is to design traffic rules in the system as in the real world. A preliminary study has been applied to MAPF by requiring agents to move in a first-come-first-served manner~\cite{Jansen2008DirectionMF}. Future work can focus on developing more systematic rules to guide the movement of agents.


\noindent \textbf{Data-Driven Future Prediction.}
Previous works have explored using data-driven methods to predict future congestion and incorporate the prediction in planning~\cite{Yu2023}, but it is only applied to simple reactive planning algorithms with $A^*$ search. Future works can explore incorporating predicted traffic information into more recent LMAPF algorithms. Furthermore, we can also consider future congestion when assigning tasks to agents.

\noindent \textbf{Applying Real-time Search.}
The key insight of real-time search is that the agent can escape the local minima by updating the heuristic~\cite{realtimeFundamental1990} during execution with a constant amount of search around the space of this agent.
The heuristic penalizes previously visited states and encourages the agent to explore other areas of the map. In the 2D MAPF context, we have a perfect single-agent Backward Dijkstra heuristic, so agents getting stuck is not due to obstacles but instead, congestion. Thus, updating agents' heuristics based on congestion could mitigate congestion caused by limited-horizon planning. Initial attempts \cite{realtimeMAPF2018} have not dealt with inter-agent interactions, but this area is ripe for advancement.

\noindent \textbf{Searching for the Optimal Number of Agents.}
The fact that disabling agents improves throughput indicates that we can search for an optimal number of agents for a given LMAPF instance. As shown in \Cref{fig:num_agents_game}, 2,500 agents are enough to achieve the best throughput.
Future work could attempt to determine the optimal number of agents to maximize the throughput. More interestingly, we can identify such a number similarly for a local region to more formally study the concept of congestion.

\section{Challenge 3: Gaps between LAMPF Models and Real-World Applications} \label{section:simtoreal}

In this section, we discuss the third challenge of \emph{bridging the gaps between commonly used LMAPF models in the literature and real-world LMAPF applications.}
We conjecture that this challenge can also deteriorate the former two challenges. Therefore, we should be aware of the existence and influence of these gaps and study more realistic LMAPF models in future research. 

\subsection{Dealing with the LMAPF-R Model}

Rather than using a four-way movement action model, which allows an agent to move to one of its neighboring vertices on a four-neighbor grid map, the competition adopts the more realistic LMAPF-R model, where the motion of agents is based on moving forward, rotating, and waiting, to close the gap between our study and the real world. There are two possible ways to approach rotations in LMAPF-R. One is to plan with the original action model without rotations and then map the solution to the new one with rotations, for example, using global synchronization or temporal plan graphs~\cite{HoenigICAPS16}. Another way is to directly include rotations in the planning algorithms. 

We adopt the second solution because direct consideration of rotation in the planning algorithms could lead to better throughput~\cite{varambally2022mapf}. Adapting LNS for rotations is simple because we only need to modify the single-agent planner used in the prioritized replanning. But it is tricky for PIBT if we want our search to be based on rotations. Because PIBT needs to reason collisions based on the next locations of agents, while rotations do not directly imply the next locations. For example, if an agent rotates 90 degrees clockwise, it is not immediately clear if the agent wants to move right or backward. Without clear information about the next locations, PIBT cannot determine if a collision will happen in the next step. So, in our solution, we still resolve collisions at each step of the search, by planning the next location first with the original PIBT. Since we have rotations, moving to the next location may correspond to multiple actions and different agents may need different steps to their next location. We ask an agent to take only the first action that leads to its next location and discard the remaining ones. In this way, our search incorporates rotations directly in the current step (as the first actions to the next locations). A recent paper \cite{zhang2023efficient} also studies the difference between the two action models and shows how to adapt CBS to the rotation action model. 

The rotation action model could more easily lead to congestion because, intuitively, it takes more steps for agents to avoid collisions when agents need to change moving directions. To verify this, we test the four-way movement action model on Random 800, Game 4000, Warehouse 8000, and Sortation 10000, namely the 4 instances on which WPPL underperforms PIBT without guidance graphs in~\Cref{tab:guidance}. We observe that WPPL performs much better than PIBT in Game 4000, Warehouse 8000, and Sortation 10000 with relative throughput improvements of 185.0\%, 192.5\%, and 189.0\%, respectively. For Random 800 with an extreme agent density, WPPL still performs worse, but only with a gap of 11. 0\%, compared to the original 22.5\%. Such results demonstrate that the rotation action model is more prone to congestion than the four-way movement action model.

Thus, we argue that even trying to close a small gap, such as adding rotation actions, still requires a certain amount of reasoning and engineering. It may also lead to different results in the experiment and requires careful investigation.

\subsection{Unrealistic Modeling}
There are several other gaps beyond the competition, which we believe are worth further studying. These gaps may be studied for MAPF, but are less explored in LMAPF. 

\noindent \textbf{Planning with Kinematics and Dynamics}. 
Even though this competition adopts a more realistic kinematic model with rotations as separate actions, it is still far simpler than many real-world scenarios. Complex kinematics and dynamics inevitably require more computational time, even for the motion planning of a single agent. In addition, complex kinematics and dynamics could also lead to more severe traffic congestion or deadlocks, since avoiding a collision involves more complex movement.

\noindent \textbf{Uncertainty during Execution}.
We also need to be aware of the uncertainty during execution. Both planning with uncertainty and replanning frequently would require large computations. Adopting an execution framework like Temporal Planning Graph (TPG) \cite{HoenigICAPS16} might be a favorable solution when facing limited planning time, but at the potential cost of less efficient execution due to the overly constrained coordination.

\noindent \textbf{Non-Uniform Task Distribution}.
We usually assume tasks are uniformly sampled from certain locations or even all traversable locations on maps in the literature. However, in LMAPF, imbalanced task distribution could lead to very different local agent densities in different regions, which may need to be addressed by different algorithms or allocated different amounts of computational resources.

\noindent \textbf{Evolving Systems}. 
We often overlook the system evolution in some applications, for example, the dynamic changing of map layouts or agent numbers~\cite{morag2022online}. 
For example, roads might be shut down due to accidents or maintenance in a traffic system; vehicles move in and out of a road intersection. These dynamic changes also require us to design more adaptive algorithms for different situations.

\section{Conclusion} \label{section:conclusion}

With the growing number of real-world LMAPF applications, it is natural to consider scaling LMAPF algorithms to more challenging problem instances. We outline in this paper three challenges identified during our participation in the 2023 League of Robot Runners competition. Firstly, we address the limited planning time constraint, proposing WPPL as a solution and suggesting future directions in rule-based, anytime, and parallel algorithms. Secondly, we alleviate the congestion issue by applying guidance graphs and disabling agents. We propose future directions, such as designing better guidance and traffic rules, incorporating future traffic prediction and real-time search, and determining the optimal agent quantity. Last, we discuss the need to reconcile the gap between simplified LMAPF models in the literature and the more complex real-world applications, considering kinodynamic models, execution uncertainty, and evolving systems.

\section*{Acknowledgements}
The work was supported by NSF Grant 2328671.

\bibliography{aaai24}

\clearpage

\end{document}